\documentclass[twocolumn,aps,10pt,nofootinbib, longbibliography,superscriptaddress,tightenlines,notitlepage,pra,floatfix,raggedbottom]{revtex4-2}
\usepackage[utf8]{inputenc}
\usepackage{physics}
\usepackage{amsfonts}
\usepackage{graphicx}
\usepackage[dvipsnames]{xcolor}
\usepackage{algorithm}
\usepackage{algpseudocode}
\usepackage{amsthm}
\usepackage{amsmath}
\usepackage{amssymb}
\usepackage{blkarray}
\usepackage{mathtools}
\usepackage{verbatim}
\usepackage{mathbbol}

\newcommand{\braid}{\text{\footnotesize $BRAID_4$}}
\newcommand{\braidtwo}{\text{\footnotesize $BRAID_2$}}

\usepackage{hyperref}
\hypersetup{
    colorlinks=true,
    citecolor=red
}
\begin{abstract}
To implement a quantum error correction protocol, we first need a scheme to prepare our state in the correct subspace of the code, and this can be done using a unitary encoding circuit. Majorana codes are special since any gates that transform such codes must preserve fermionic parity. In this paper, we present an algorithm that uses the stabilizer matrix to compute unitary encoding circuits for Majorana codes. We present two approaches, both of which use a version of Gaussian elimination with row operations replaced with elementary fermionic Clifford operations. One approach uses an additional ancilla mode and works for all Majorana stabilizer codes, while the second approach does not use ancilla but does not work if the total parity is inside the stabilizer group. 
\end{abstract}
\begin{document}

\title{Encoding Majorana codes}
\author{Maryam Mudassar$^*$}
\affiliation{Joint Center for Quantum Information and Computer Science, University of Maryland, College Park 20742, USA}
\affiliation{Department of Physics and Astronomy, Dartmouth College, Hanover, NH 03755, USA}
\author{Riley W. Chien$^{*\dag}$}
\affiliation{Department of Physics and Astronomy, Dartmouth College, Hanover, NH 03755, USA}
\def\thefootnote{$^*$}\footnotetext{\href{maryammu@umd.edu}{maryammu@umd.edu}}
\def\thefootnote{$^\dag$}\footnotetext{\href{rileywchien.gr@dartmouth.edu}{rileywchien@gmail.com}}
\author{Daniel Gottesman$^{**}$}
\affiliation{Joint Center for Quantum Information and Computer Science, University of Maryland, College Park 20742, USA}
\def\thefootnote{$^{**}$}\footnotetext{\href{dgottesm@umd.edu}{dgottesm@umd.edu}}
\maketitle

\section{Introduction}
Quantum computation is highly prone to errors, and many efforts have been made to make computation fault tolerant, and the central principle of fault tolerance relies on error correction. However, a major challenge to fault tolerance remains the fact that error rates for most platforms \cite{xu:2023,acharya2022suppressing,IBM_future_2022} are higher than the error threshold for quantum error correction. Amongst different models of quantum computers, topological quantum computing is a viable contender since qubits are encoded in non local degrees of freedom \cite{RevModPhys.80.1083}, which protects them against local errors. In a Majorana based hardware, logical gates are executed by braiding anyons, which can be performing by physical braiding or via measurement \cite{Tran_2020}. In order to have access to a universal gateset, one must have a sufficiently rich anyon theory. While Majoranas do not accommodate a universal gateset since they only realize the Clifford group, they are an attractive candidate for realizing topological quantum computation, in part due to the clear path to realizing Majorana zero modes in mesoscopic devices.  Such modes are believed to be hosted at the interface of semiconductor and p-wave superconductors \cite{PhysRevB.95.235305, PhysRevLett.105.077001, Alicea_2012, Aghaee_2023}, and qubits can be encoded into the degenerate subspace of the Majorana edge modes.

This can be recast in the language of error correction, and these qubits can be understood to be the logical qubits encoded in the physical Hilbert space of Majoranas. These qubits are both protected against local perturbations, as well as parity switching errors due to fermionic superselection \cite{Bravyi_2010, Kundu_2023, hastings2017smallmajoranafermioncodes}. Additionally, it is possible to measure qubits in all three bases with limited decoherence allowing for the implementation of robust single qubit Clifford gates with zero time overhead \cite{Litinski_2018}. Another motivation to consider Majorana codes is they specify a bigger class of error correcting codes than qubit stabilizer codes. It has been shown that an $[[n,k,d]]$ qubit stabilizer code can be mapped to a $[[4n,k,2d]]$ Majorana code, but this requires adding local parity to the Majorana stabilizer \cite{Bravyi_2010}. However, an arbitrary Majorana code which may not have this parity constraint cannot be converted back to a qubit stabilizer code with the same number of logical qubits. Secondly, as stated in \cite{mclauchlan2023new}, Majorana surface codes can encode both fermionic and bosonic twists, which is novel compared to qubit based surface codes that can encode only fermionic twists at the corners \cite{landahl2023logicalfermionsfaulttolerantquantum}. Last but not the least, for native fermionic hardware, it can be useful to have codes that correct for fermionic noise, and we expect that fermionic codes would also find use in such platforms \cite{Gonz_lez_Cuadra_2023}.

To implement these Majorana codes, we need to find a way to encode the logical state. The current dominant approach is to measure the stabilizers, and this requires doing 2-Majorana zero mode (MZM) and 4-MZM measurements and entangling gates with ancillary qubits. \cite{Paetznick_2023, Tran_2020}). However, given the difficulties to date to realize Majoranas \cite{zhang2021retraction, he2017chiral}, it is beneficial to have multiple options to encode a logical state. To this end, we propose unitary encoding circuits for Majoranas that use braiding operations. For qubit stabilizer codes, it is natural to use the method in \cite{Cleve_1997} to construct encoding circuits, that map an initial product state on $k$ qubits and $n-k$ ancilla to an entangled state on $n$ physical qubits. We use a similar approach using the binary vector formalism for Majoranas.

 In this paper, we propose a method to find encoding circuits for a general Majorana stabilizer code. The circuit is composed of Majorana quadratic and quartic braid gates, which belong to the Majorana Clifford group since they normalize the Majorana stabilizer group. The algorithm has two variants, one which uses two ancilla modes (which are reset at the end of the circuit). In the second algorithm, we do not use any ancilla modes but if the total parity is inside the stabilizer group then this algorithm fails. Note that our construction of the encoder here is not fault tolerant. In the non-fault tolerant case, the code and construction of the encoder  can still be useful for storage of information for long periods and communication over a noisy channel.  We give a proof of the two algorithms in Appendix A, and also a short proof of why the second algorithm fails in Appendix C.  
\section{Background}
\subsection{Majorana codes}
In 2010, Bravyi et al \cite{Bravyi_2010} initiated the study of Majorana fermionic codes as extensions of the 1D Kitaev chain \cite{Kitaev_2001}. These Majorana codes can be used to protect against low weight fermionic errors, which is suitable for Majorana-based platforms that host Majorana zero modes, such as semiconductor-superconductor heterostructures \cite{Alicea_2012}. These Majorana codes provide topological protection to an encoded qubit, as well as give a construction that maps between logical qubits and physical Majorana degrees of freedom. One unique feature of these fermionic codes is the presence of superselection rules, which is the source of protection in certain codes with even logical operators. 

A set of $n$ complex fermions are described by fermionic creation and annihilation operators with the following anticommutation relations:
\begin{align}
\{a_i,a_j^{\dagger}\}=\delta_{i,j}
\end{align}
To a set of $n$ pairs of creation and annihilation operators, we can associate $2n$ Majorana operators: 
\begin{align}
c_{2i} = a_i + a_i^{\dagger} \qquad c_{2i+1} = i(a_i - a_i^{\dagger})
\end{align}
which have the following relations:
\begin{align}
\{c_i,c_j\} = 2\delta_{i,j}, \quad c_i^2=\mathbb{1} 
\end{align}

These Majorana operators, together with the phase factor $i$, generate the non Abelian group $\mathrm{Maj}(2n)$ \cite{bettaque2024structuremajoranacliffordgroup}, hence any Majorana operator can be written as a product of single mode operators $c_i$ and a phase factor $\eta\in\{\pm 1,\pm i\}$. The weight of a Majorana operator is defined as the number of Majorana modes in its support. The individual fermion parity $P_j$ is mapped to $ic_{2j}c_{2j+1}$, and the total parity operator is $P_{tot} = \prod_{j=1}^{N-1} i c_{2j}c_{2j+1}$. 
Due to the fermion parity superselection rule, we cannot have a coherent superposition of even and odd parity states. Any projective measurements, for instance stabilizer measurements, must therefore comprise of operators of an even weight, and thus commute with the total parity. On the other hand, logical operators are allowed to have odd weight, for instance in the case of quasiparticle poisoning when a single electron tunnels between states that hold information and the environment, one has single Majoranas locally which changes the parity of the encoded qubit \cite{https://doi.org/10.48550/arxiv.1703.00459}. 

The Majorana stabilizer group $\mathrm{S_{Maj}}$ is an Abelian subgroup of the $\mathrm{Maj(2n)}$ group. The generators have even weight, square to $+\mathbb{1}$ and do not contain $-\mathbb{1}$. The codespace of $\mathrm{S_{Maj}}$ corresponds to the ground subspace of a Hamiltonian $H=-\sum_{i} S_i$ that may involve geometrically local or non local interaction between the modes, in general, and while the encoding algorithm works in general for all Majorana codes, here we have specifically shown examples of 1D and 2D local topological codes.

 A Majorana code with  $n$ complex fermions or $N=2n$ Majorana fermions that encodes $k$ logical qubits and has code distance $d$ is denoted as a $[[n,k,d]]_f$ code. Such a code has $r=n-k$ independent stabilizer generators and encodes $k$ qubits in the $2^k$ degenerate subspace of the stabilizers. 

The set of Majorana operators that commute with the stabilizer is called the \textit{normalizer} of $S_\mathrm{{Maj}}$ and is denoted as $N(S_\mathrm{{Maj}})$. The non trivial logical operators of the Majorana group are described as elements that commute with $\mathrm{S_{Maj}}$, but are not inside the stabilizer i.e. $L\in N(S_\mathrm{{Maj})}$\textbackslash $S_\mathrm{{Maj}}$.  As discussed before, it is  possible to have codes in which logical operators have odd weight, and the Majorana color code is one example of such a code. 

The code distance $d$ is the minimum number of Majorana operators in a logical operator.  However, note that in general, code distance does not completely capture the stability offered by a Majorana code. An additional parameter, $l_{even}$ is sometimes introduced, as the minimum diameter of a region that can support an \textit{even} logical operator. More formally, we restate the definition of $l_{even}$ given in \cite{Bravyi_2010}:
\begin{align}
l_{even}= \min_{C\in \mathcal{C}(S)\backslash S |C|=0 \pmod 2} \text{diam}(\text{Supp}(C))
\end{align}Such an operator corresponds to errors when the underlying system is considered to be closed, or interacting with a bosonic environment. Normally, code distance measures the topological robustness of code, while $l_{even}$ gives the protection by superselection rules. 
Below are a few examples of well-known Majorana codes:
\subsection{Example: 1D Kitaev chain}
Starting from the following Hamiltonian,
\begin{align}
H = -\mu \sum_{i} a_i^{\dagger}a_i -t\sum_i (a_{i+1}^{\dagger}a_i + h.c.) + \notag \\
\Delta \sum_i (a_i a_{i+1} + h.c.).
\end{align}
We take the limit $\mu=0, \Delta=t$, and rewrite fermionic operators in terms of Majorana operators.
\begin{align}
    H = -i\Delta \sum_{i=1}^{N-1} c_{2i}c_{2i+1}. 
\end{align}
For $\Delta = 1$, the Kitaev chain can be used as a Majorana code if we define the stabilizer of the code to be:
\begin{align}
S_\mathrm{{Maj}}=\langle ic_{2i}c_{2i+1}\rangle.
\end{align}
while the logical operators are $\bar{X_1}=c_1,\bar{Z_1} = c_{2n} $. Superselection rules forbid unpaired Majorana operators to act, hence the only logical errors are $c_1c_{2n}$, and thus this code protects the encoded qubit against any local even parity fermionic errors. The 1D Kitaev chain is $[[n,1,1]]_f$ but $l_{even}=2n$, which protects the code against low weight even fermionic errors but still exposes it to quasiparticle poisoning.

\subsection{Example: The shortest fermion code}
Another example to note is the shortest fermion code or the $[[6,1,3]]_f$ code which was introduced in \cite{https://doi.org/10.48550/arxiv.1703.00459}, which encodes a single logical qubit and corrects for all elementary fermion parity flip errors, and is the shortest fermion code to do so. The stabilizer is given as: 
\begin{align}
O_1 &= c_1c_2c_3c_4\notag\\
O_2 &= c_3c_4 c_5 c_6\notag\\
O_3 &= c_7c_8c_9c_{10}\notag\\
O_4 &= c_9 c_{10} c_{11} c_{12}\notag\\
O_5 &= ic_2c_4c_6c_8c_{10}c_{12}
\end{align}
while the logical operators are given as:
\begin{align}
\Gamma_1^x &= c_1c_3c_5\notag\\
\Gamma_1^z &= c_2c_4c_6c_7c_8c_9c_{10}c_{11}c_{12}
\end{align}
such that the product of the two logical operators is equal to the total fermion parity upto a phase. They also protect against dephasing errors like $c_{2i}c_{2i+1}$ since each parity term anticommutes with each of the stabilizer generators. One difference to note with respect to Kitaev chains is that these codes are capable of correcting all elementary weight-1 errors, or parity flip errors.  
\subsection{Example: Majorana color codes}
Majorana color codes \cite{Bravyi_2010, Litinski2018colorcode,Vijay_2015} describe a fermionic version of topological color codes introduced by Bombin et al \cite{BombinColorCode}. To define a Majorana color code, we choose a graph G that describes a tesselation of a surface, and the graph must be tri-colorable (the codes mentioned here are 2D Majorana color codes, in general, one can generalize this to the cellulation of a  $D-$dimensional manifold, where each the vertex is $D+1-$valent and the graph is $D+1$ colorable \cite{Kubica_2018}). Such a graph has an even number of vertices $V$, where each vertex corresponds to a Majorana mode, and the product of modes around a face corresponds to a stabilizer. Such a graph $G$ can be embedded into a two dimensional cylinder $\Sigma$. 
The logical operators for these codes are of odd weight, and the main advantage of such codes is that the encoded qubit has an additional degree of protection in the size of the cylinder, and as with all topological codes, we can control the distance by choosing the geometry of the code. A simple example of a Majorana color code is the 2D color code on the hexagonal lattice with periodic boundary conditions in the horizontal direction.  \cite{Bravyi_2010}. The stabilizers for the code are given as follows, for a hexagonal graph:
\begin{figure}[h]
\includegraphics{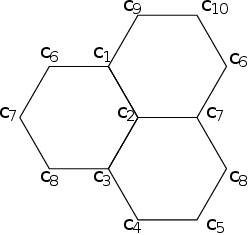}
\caption{Majorana color code on ten modes. }
\end{figure}
\begin{align}
O_1 & = c_1 c_2 c_3 c_6 c_7 c_8\\
O_2 & = c_1 c_2 c_6 c_7 c_9 c_{10}\\
O_3 &= c_2 c_3 c_4 c_5 c_7 c_8
\end{align}
One can obtain the following choice of logicals:
\begin{align}
\Gamma_1^{x}&= c_1 c_2 c_3 c_6 c_9\\
\Gamma_1^{z}&= c_9 c_{10}\\
\Gamma_2^{x}&= c_1 c_2 c_5 c_9 c_{10}\\
\Gamma_2^{z}&=  c_1 c_3 c_6 c_8 c_9 c_{10}
\end{align}

\section{Clifford group for Majoranas}
The Gottesman-Knill theorem \cite{gottesman1998heisenberg} states that a stabilizer circuit $C$ acting on $n$ qubits, that is a quantum circuit comprised of Clifford operations, maps any element inside the Pauli group $P_n$ to any other element $P_n$ under conjugation.   
The binary matrix formalism is useful here, since Pauli strings can be written as:
\begin{align}
P = i^r X_1^{a_1} X_2^{a_2}\cdots X_n^{a_n}Z_1^{a_{n+1}}Z_2^{a_{n+2}}\cdots Z_n^
{a_{2n}} \notag\\ r\in \{0,1,2,3\}, a_i \in \{0,1\}.
\end{align}
and hence can be represented, modulo phases, as binary vectors of length $2n$. The Clifford group elements correspond to symplectic matrices over the $2n$ dimensional vector space, such that they map Pauli elements to themselves and preserve the symplectic form \cite{gottesman2009introduction}. 

We are interested in an analogous construction for Majorana operators, with Clifford operations mapping Majorana monomials to themselves, and that they preserve the fermionic symplectic form.  Note that we can define any element in the Majorana group $\mathrm{Maj(2n)}$ as a Majorana monomial:
\begin{align}
M&= i^{r} c_1^{m_1}c_2^{m_2}\ldots c_{2n}^{m_{2n}}, r\in \{0,1,2,3\}, m_i \in \{0,1\}\notag\\
M&\coloneqq i^r [\bar{m}].
\end{align}
where $[\bar{m}]$ is the Majorana operator mod phase, where by convention, the product is taken in the order shown, from lowest index to highest. We can represent this as a binary vector $\bar{m}$ of size $2n$:
\begin{align}
\bar{m}^T = [m_1,m_2,\ldots m_{2n}]\qquad m_i \in \{0,1\}
\label{majorana_binary_vector}
\end{align}
The Majorana Clifford group is defined to be the group of unitaries that by conjugation, map Majorana monomials to Majorana monomials. This group of gates can be shown to be equal to the group of gates consisting of $\frac{\pi}{4}$ rotations generated by even parity Majorana monomials.
Moreover, all such rotations can be generated by a quartic Majorana operator, which we refer to as $\braid$, and has been introduced in \cite{McLauchlan_2022}: 
\begin{align}
\braid(a,b,c,d)= \exp(i\frac{\pi}{4} c_a c_b c_c c_d)
\end{align}
The reason for calling this gate a braiding gate is because it braids logical Majoranas together, if we consider $c_a$ and $c_bc_cc_d$ to be logical Majorana degrees of freedom \cite{mclauchlan2023new}.
 More explicitly, the action of  $\braid(\bar{c})=\exp(i\frac{\pi}{4}[\bar{c}])$ is given as:
\begin{align}
\braid(\Bar{v})[\bar{m}]\braid(\Bar{v})^\dagger) =\frac{i}{2}(1-(-1)^{|\Bar{v}^{T} \Lambda_f \Bar{m}|}) [\Bar{v}][\Bar{m}] +\notag\\ \frac{1}{2}(1+(-1)^{|\Bar{v}^{T}\Lambda_f \bar{m}|})[\Bar{m}].
\label{eq:braidaction}
\end{align}
where $\Lambda_f$ is the fermionic symplectic form \cite{chien2022optimizing} defined as:
\begin{align}
\Lambda_f = \mathbb{1}+C
\end{align}
where $C$ is a constant matrix with 1 at every entry. Note that the output depends on $|\bar{v} \Lambda_f \bar{m}|$, which gives the overlap between $[\bar{v}]$ and $[\bar{m}]$, which can be even or odd. $\braid(\Bar{v})$ acts as identity if $|\bar{v} \Lambda_f \bar{m}|$ is even, and as $\bar{v}\bar{m}$ if it is odd. In particular, if a $\braid(\bar{v})$ has three (one) overlapping modes with $[\bar{m}]$, then it reduces (increases) the weight of the binary string $\bar{m}$ by two. This observation is important when we choose gates for our decoding algorithm.   
We can also find the binary matrix form of the $\braid(\Bar{v})$ gate, given in the Appendix, which allows us to verify 
\begin{align}
\braid(\bar{m})\Lambda_f (\braid(\bar{m}))^{\dagger}=\Lambda_f 
\label{symplectic_condition}
\end{align}
which indeed fulfills both conditions of being a Clifford operator, since they map Majorana operators to themselves under conjugation and preserve the fermionic symplectic form.

In addition to updating the phaseless portion of a Majorana monomial specified by the bitstring, the action of a gate will generally update the phase as well. With the product of Majoranas taken in the conventional order, the phase of each generator is specified by a $\mathbb{Z}_4$ variable, $r$, that determines the integer power of $i$. When the gate $\exp(i\frac{\pi}{4}V)$ acts on M, then the phase variable $r$ is updated as:
\begin{align}
    r \to r + |\bar{v}^T\Lambda_f \bar{m}|(1+r' + 2\bar{v}^T \Lambda_f^L \bar{m}).
\end{align}
where $V=i^{r'} [\bar{v}]$, and $|..|$ is the overlap of $[\bar{v}]$ and $\bar[{m}]$ modulo two.

In the case of $\braid(\Bar{v})$, this becomes: 
\begin{align}
    r\to r + |\bar{v}^T \Lambda_f \bar{m}|(1+ 2\bar{v}^T \Lambda_f^L \bar{m}).
\end{align}
where $\Lambda_f^L$ is the lower triangular portion of $\Lambda_f$. The first term in the update comes from the factor of $i$ in the update in  (\ref{eq:braidaction}). The second term is the phase picked up by commuting Majoranas when taking the product $[\bar{v}][\bar{m}]$ to place Majoranas into the conventional ordering. The action of $\braidtwo$ on the phase is in the Appendix. 

We also use the $\braidtwo$ gate extensively in the algorithms below, and both its action and its binary form is given in the Appendix. Note that a product of $\braid$ gates also generate the same action as a $\braidtwo$ gate on a Majorana string, but this does not mean that a $\braidtwo$ is a product of $\braid$ gates. For example, for $\bar{m}=c_1c_2$, conjugating by $\exp(i\frac{\pi}{4} c_1c_3c_4c_5)\exp(i\frac{\pi}{4} c_1c_2 c_4 c_5 )$ has the same action as conjugating by $\exp(-\frac{\pi}{4} c_1c_3)$, up to phases. 

\section{Encoding codewords}

In order to use a code to protect and process quantum information, we require a way of instantiating the system into a desired codeword.
Here, we are interested in a unitary encoding circuit.

Given a binary representation of the stabilizer generators, one can find a quantum circuit $C$ \cite{gottesman1997stabilizer} for qubit stabilizer codes that finds that maps basis states to codewords $\ket{\Psi}$:
\begin{align}
C\ket{0,0,0\ldots,0,c_1,c_2,..c_k}\rightarrow \ket{\psi}
\end{align}
where $\ket{\psi}$ is an encoded codeword over $n$ qubits. This is equivalent to mapping the trivial stabilizer $S_q^d$ which corresponds to single qubit Z's to  the encoded stabilizer $S_q^e$.  

The problem this paper aims to address is how to perform an encoding for a Majorana code, by mapping the trivial Majorana stabilizer $S_{Maj}^d$, which corresponds to single mode parities $c_{2i}c_{2i+1}$, to the stabilizer for the Majorana code $S_{Maj}^e$. 
Note that one approach to encode such codes would be by using the qubit to fermion mapping defined in \cite{Bravyi_2010}, however since this approach includes individual parity inside the stabilizer, and as a consequence the total parity is also included inside the stabilizer, this method will only work for a restricted class of fermionic codes i.e. codes with even weight logical operators. Our method is general for the entire class of Majorana stabilizer codes.

We do this by giving the algorithm for decoding which maps the given stabilizer for the Majorana code to the trivial stabilizer, and since this circuit is unitary, it can easily be inverted to get the encoding circuit. Additionally, using the encoding circuit our algorithm produces, we can easily compute destabilizers which could have use in classical simulation.

\section{Algorithm for encoding: with ancilla}
We now outline the steps to finding the decoding circuit $C$ consisting of the above $\braid$ and $\braidtwo$ gates that maps the encoded stabilizer $S^{e}$ to the decoded stabilizer $S^d$. We can represent the stabilizer generators of the Majorana code as binary vectors using the expression in Eq (\ref{majorana_binary_vector}), where $r$ is the number of generators and $N=2n$ is the number of Majorana fermions. The binary matrices for the stabilizers are: 
\begin{align}
S^{e} = \begin{bmatrix}
s_{11} & s_{12} & \cdots s_{1r}\\
s_{21} & s_{22} & \cdots s_{2r}\\
\vdots & \vdots & \vdots \\
s_{N1} & s_{N2}&  \cdots s_{Nr}
\end{bmatrix}
\end{align}

\begin{align}
S^{d}=\begin{bmatrix}
1 & 0 & 0 & 0  \cdots 0 & 0 & 0 \\
1 & 0 & 0 & 0  \cdots 0 & 0 & 0 \\
0 & 1 & 0 & 0 \cdots  0 & 0 & 0 \\
0 & 1 & 0 & 0 \cdots  0 & 0 & 0 \\
\vdots  &\vdots& \vdots &\vdots & \vdots & \vdots\\
\end{bmatrix}
\end{align}
where the columns in each matrix represent the stabilizer generators. Note that by virtue of being a stabilizer matrix, both $S^e$ and $S^d$ have the following properties:
\begin{itemize}
\item Each column should have an even number of ones.
\item Each column must have an even number of overlapping ones with any other column.
\item The matrix must have full column rank. 
\end{itemize}

The outline of the algorithm is as follows. Note that the \textbf{pivot p} is defined as the index below which all entries (except the one right below) are zero after decoding.  
\begin{itemize}
    \item Add two extra zero rows to the start of the stabilizer matrix. These correspond to the initially unoccupied ancilla modes.
    \item Pick a column in $S_e$. Starting from the \textbf{pivot} index, apply \textbf{\braid} and $\braidtwo$ gates until all the elements below the pivot+1 index are zero. Use the ancilla mode as the unoccupied mode, and if it is not zero, swap with another unoccupied mode below the pivot index. 
    \item Use the pivot index as the ``control" mode to clean all the ones above the pivot index, use $\braidtwo$s with the ancilla to reset the pivot mode to 1, and ancilla to 0. Using this method will ensure that all the columns before the selected one do not change. 
\end{itemize}
A longer description of the algorithm is stated in Algorithm \ref{alg:decoding}. 

\begin{algorithm}[H]

\caption{Construct decoding circuit for a binary stabilizer matrix with ancilla $\hat{S}$}\label{alg:decoding}
\begin{algorithmic}[2]
\Function{DecodingCircuit}{$\hat{S}$}
\State $\hat{S} \leftarrow \{r_1,r_2\}\cup \hat{S}$
\State $p \leftarrow 2$
\For {$i$th column in $\hat{S}$}
    \State Store non-zero indices of $\hat{S}_i$ from row $p$ to $2n$ in $arr$
    \State Store phases of $\hat{S}_i$ in Phase
    \While{weight(arr) $> 0$} 
        \If{weight(arr)is $ 2$}
            \State $\hat{S} \gets \braidtwo(p,arr(1))\cdot \hat{S}$
            \State $\hat{S} \gets \braidtwo(p+1,arr(2))\cdot \hat{S}$
            \State Update phases in Phase
        
        \Else
        \State $\hat{S}  \gets \braid(0,arr(1),arr(2),arr(3))\cdot\hat{S}$    
        \State Delete arr(2) and arr(3)
        \State $\hat{S}  \gets \braidtwo(0,arr(1)) \cdot \hat{S}$
        \State Update phases in Phase
        \EndIf
    \EndWhile
    \State Store non-zero indices of $S_i$ from row 0 to p
    \While{weight(arr) $> 1$} 
        \State $\hat{S} \gets \braid(0,arr(1),arr(2),p)\cdot \hat{S} $
        \State Delete arr(1) and arr(2)
        \State $\hat{S}  \gets \braidtwo(0,p)\cdot \hat{S} $
        \State Update phases in Phase
    \EndWhile 
      
\State$p\gets p + 2$
 
\EndFor\\
\For {$k$th column in $\hat{S}$ }
    \If{Phase[k] = -i and k = 1}
    \State $\hat{S} \gets \braid(0,k,2r,2r+1)$ .$\hat{S}$
    \State $ \hat{S} \gets \braid(0,k,2r,2r+1)$ .$\hat{S}$
    \ElsIf{Phase[k] = -i}
    \State $\hat{S} \gets \braid(0,2k-1,2k-2,2k)$ .$\hat{S}$
    \State $ \hat{S} \gets \braid(0,2k-1,2k-2,2k)$ .$\hat{S}$
    \EndIf
\EndFor\\
\Return $\hat{S}$
\EndFunction
\end{algorithmic}
 
\end{algorithm}

From the relation stated in Eq. \ref{symplectic_condition}, we have the symplectic condition:
\begin{align}
    C^{T} \Lambda_f C = \Lambda_f
\end{align}
which allows us to find the inverse of decoding, corresponding to the encoding circuit:
\begin{align}
C^{-1} = \Lambda_f^{-1} C^{T}\Lambda_f 
\end{align}
The gate complexity for this algorithm is $O(r N)$ where $r$ is the number of stabilizer generators and $N$ is the number of Majorana modes.  
\section{Phase tracking and correction}
During the encoding circuit it is possible to flip the phases of the generators, and prepare the state in the syndrome space. In general it is also possible to track the phases and correct them at the end of the circuit in a parity preserving way. Using the update rule from section III, we can update $r$ $\mathbb{Z}_4$ variables for each of the $r$ generators. 

We now describe the additional steps for the decoding algorithm to do this correction. After the decoding, our stabilizer will be in the decoded form $S^{d'}$, but the signs could be flipped:
\begin{align}
S^{d'}_i = \pm i c_{2i}c_{2i+1} 
\end{align}
To fix that, we check the $\mathbb{Z}_4$ variable for each of the generators and if the phase is $-i$ for a generator, say $S_i$, then we can choose a $\braid$ gate that has weight one overlap with $S_i$, weight one overlap with a logical mode(labelled with N in Algorithm \ref{alg:decoding2}), and even overlap with all of the other generators. Even overlap with this $\braid$ will ensure that the sign of the other generators is not changed. Applying this $\braid$ twice will map $S_i$ to $-S_i$, but also flip the sign of the logical operator. Since a logical operator with a flipped phase still belongs to the same group, (e.g. $\bar{Z_k}\bar{X_k}\bar{Z_k}=-\bar{X_k}$), this operation is valid and maps it to the correct subspace of the stabilizer code. If an even number of logical generators have the wrong sign, then a similar strategy can be used to choose a $\braid$ that flips their signs, however in the event that there are odd number of generators, one must use ancilla to flip the logical operators in a parity preserving manner. 
\section{Examples of encoding circuits}
\begin{figure}
 \includegraphics[]{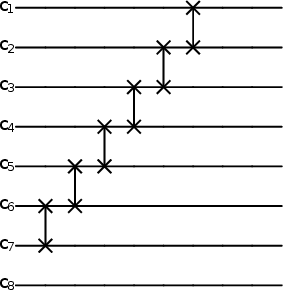}
\caption{Encoding circuit for 1D Kitaev chain. Here the gates are just the $\braidtwo$ gates.}
\label{fig:figure_1}
\end{figure}

We now give a few examples of the encoding circuits for known Majorana codes using the construction given above. In Figure \ref{fig:figure_1}, we have the encoding circuit for a 1D Kitaev chain.

\begin{figure}
\includegraphics[]{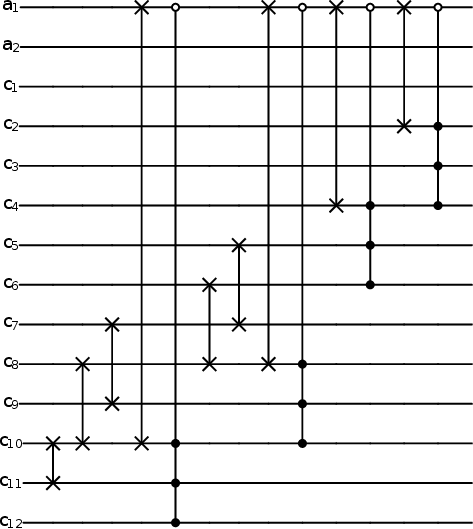}
\caption{Encoding circuit for $[[6,1,3]]_f$. Note that the only non-local gates are acting on the ancilla mode, labelled as $a_1$ and $a_2$.  Here the two mode gate is the $\braidtwo$ gate, and the four mode gate is the \braid gate, with the empty circle denoting the "zeroMode" as described in the algorithm.}
\label{fig:my_label2}
\end{figure}

The encoding circuit for the shortest Majorana code that corrects for parity flip errors is given in Figure \ref{fig:my_label2}. Note that with the construction using ancillas, our gates are no longer acting geometrically locally. 

In the next section, we give a modified implementation of the algorithm, which is ancilla-free, and use the ancilla free implementation for the encoding circuit of a Majorana color code. 

\section{Ancilla free implementation}
In the above implementation, we use ancilla modes as the unoccupied mode to choose an appropriate \textbf{\braid} operation. In some cases, however, it is possible to have an ancilla free implementation, particularly when the stabilizer matrix is sparse. The outline of this modified algorithm is as follows:
\begin{itemize}
    \item Pick a column in $S_e$. Starting from the pivot index, apply $\braid$ and $\braidtwo$ gates until all the elements below the pivot+1 index are zero, and there is 1 at the pivot and pivot+1 positions. Use one of the unoccupied modes \textbf{below} the pivot index for the $\braid$ gate.
    \item Use the pivot index as the ``control" mode to clean all the ones before, and use $\braidtwo$s with pivot+2 to reset the pivot to 1. 
\end{itemize}
Algorithm \ref{alg:decoding2} gives the pseudocode for this implementation. 

\begin{figure}[h!]
\begin{algorithm}[H]
\caption{Construct an ancilla free decoding circuit}\label{alg:decoding2}
\begin{algorithmic}[2]
\Function{AncillaFreeDecodingCircuit}{$\hat{S}$}
\State $p \gets 2$;
\For{ith column in $\hat{S}$}
\State Store non-zero indices from row $p$ to $2n$ in arr
\State Store zero index from row $p$ to $2n$ in $c_0$
\State Store phases of $\hat{S}_i$ in Phase
\While{weight(arr)$ > 0$}
\If{weight(IndArr)is $ 2$}
 \State $\hat{S} \gets \braidtwo(p,arr(1))\cdot \hat{S}$
 \State $\hat{S} \gets \braidtwo(p+1,arr(2))\cdot \hat{S}$
 \State Delete corresponding indices from arr
 \State Updates phases in Phase
\Else
   \State $\hat{S} \gets \braid(c_0,arr(1),arr(2),arr(3))\cdot \hat{S}$ 
    \State Delete corresponding indices from arr
    \State Update phases in Phase
\EndIf
\EndWhile
\State arr $\gets g_i.index(1,0,p)$
\State $c_0 \gets p + 2$ 
\While{$size(\text{arr})>0$ }
\State $\hat{S}\gets \braid(c_0,p,arr(1),arr(2))\cdot \hat{S}$
\State $\braidtwo(c_0,p)$
\State Delete corresponding indices from arr
\State Update phases in Phase
\EndWhile
\State $p \gets p + 2$
\EndFor\\
\For {$k$th column in $\hat{S}$ }
    \If{Phase[k] = -i and k = 1}
    \State $\hat{S} \gets \braid(0,2k,2r,2r+1)$ .$\hat{S}$
    \State $ \hat{S} \gets \braid(0,2k,2r,2r+1)$ .$\hat{S}$
    \ElsIf{Phase[k] = -i}
    \State $\hat{S} \gets \braid(2k-1,2k-2,2k,N)$ .$\hat{S}$
    \State $ \hat{S} \gets \braid(2k-1,2k-2,2k,N)$ .$\hat{S}$
    \EndIf
\EndFor\\ 
\Return $\hat{S}$
\EndFunction
\end{algorithmic}
\end{algorithm}
\end{figure}
\begin{figure}[H]
\centering
\includegraphics[width=0.3\textwidth]{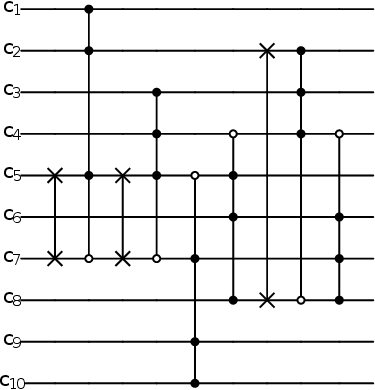}
\caption{Encoding circuit for Majorana color code on ten Majorana modes. Here the two mode gate is the $\braidtwo$ gate, and the four mode gate is the $\braid$ gate, with the unfilled circle denoting the "zeroMode" as described in the algorithm. }
\label{fig:figure_3}
\end{figure}

We can use the above method to find the encoding circuit for a Majorana color code on ten modes. The circuit is printed in Figure \ref{fig:figure_3}. Note that this code is different from the one adapted from qubit codes since for those codes the total parity is inside the stabilizer. 

 There is one caveat to this method: it does not work for all kinds of stabilizer codes. In particular, if the total parity is inside the stabilizer group, one cannot decode it without an ancilla. 
 \section{Conclusion}
 In this paper, we have presented a scheme that finds encodings for a Majorana code with $O(Nr)$ gates, for $N$ Majoranas and $r$ stabilizer generators, which is comparable with the method for qubit encodings. We have presented schemes to find a gate array using an ancilla, and also one that does not use an ancilla. We expect that this method will be useful in storing qubits in non-local degrees of freedom on Majorana based hardware, and in communicating qubits over a noisy fermionic channel. 
 
 \section{Acknowledgments}

 M.M. would like to thank James D. Whitfield, Brent Harrison and Muhammad Khan for a critical reading of the manuscript and for insightful discussions. R.C and M.M. would like to acknowledge funding from NSF (PHYS- 1820747) and Fundamental Algorithmic Research for Quantum Computing (FAR-QC), DE-SC0020411. M.M and D.G. also acknowledge funding from National Science Foundation (RQS QLCI grant OMA-2120757). 

\bibliographystyle{unsrt}
\bibliography{ref.bib}

%% The added code line

\appendix
\section{\\Proof that the algorithm for decoding circuit works for valid choices of stabilizer}
We want to prove by induction that one can use the decoding algorithm that corresponds to a decoding circuit $C_{2n \times 2n}$ to map the stabilizer matrix $S^{e}_{2n\times g}$ of a Majorana code to the decoded stabilizer $S^{d}_{2n \times g}$. It is important to note that for a complete set of generators, $g\leq n$. The matrix equation for the above is:
\begin{align}
C_{2n\times 2n}S^{e}_{2n\times g} = S^{d}_{2n\times g}
\end{align}
The base case is $g=1$. For a single generator, it is easy to show that one needs only \textbf{$\braid$} and \textbf{$\braidtwo$} to map a binary string of even weight to another binary string of weight two, i.e. simply apply a \textbf{$\braid$} gate with overlap of three modes repeatedly, until weight is two, and then swap the two modes to the correct position of the modes. There needs to be at least one unoccupied mode for this to work, otherwise one can use an ancilla. 
Now we do the inductive step i.e. if the above is true for $g=k$, it must also be true for $g=k+1$. More precisely, if we have:
\begin{align}
C_{2n\times 2n}^{(k)}S^{e}_{2n\times k} = S^{d}_{2n\times k}
\end{align}
where $S^e_{2n\times k}$ is the encoded stabilizer with $k$ columns, and $S^d_{2n\times k}$ is the decoded stabilizer with $k$ columns.
We want to show that:
\begin{align}
C^{(k+1)}_{2n\times 2n}S^{e}_{2n\times k+1} = S^{d}_{2n\times k+1}
\end{align}
Define $C^{(k+1)}_{2n\times 2n} = L_{2n\times2n}C^{(k)}_{2n\times 2n}$. Plugging in the equation for $g=k+1$,
\begin{align}
C^{(k+1)}_{2n\times 2n}S^{e}_{2n\times k+1}&= L_{2n\times2n}C^{(k)}_{2n\times 2n} S^{e}_{2n\times k+1}\\
&= L_{2n\times2n}  S'_{2n\times k+1}
\end{align}
where in $S'_{2n\times k+1}$, the first $k$ columns correspond to $S^d_{2n\times k}$, while the last is a binary string of even weight. We need to choose an $L_{2n\times 2n}$ such that it acts as identity on the first $k$ columns, but changes the last column to the decoded generator. For the $k+1$ column, we split into two subcolumns $A$ and $B$, and call the $2k+1$ row the \textbf{pivot} row. The $k+1$ column is given as:
\begin{align}
s_{k+1} = \begin{bmatrix}
    A_{2k\times1}\\
    B_{2n-2k+1\times 1}
\end{bmatrix}
\end{align}
Entries in $A$ are such that either elements at row $2i-1$ and $2i$ for $i=1,\ldots k$ are both 0 or they are both 1, so that they commute with earlier columns, whereas elements in $B$ can be any random combination of 0's and 1's, but the weight of $B$ is even.  Let the bitstring corresponding to the column be $\Bar{m}$. We now define a \textbf{shrink-and-swap} sequence to reach the desired stabilizer form. In the \textbf{shrink} part, one chooses $\braid(\Bar{v})$ gates that shrink the column weight by two. For this step, we choose a $\Bar{v}$ such that the number of overlapping modes between $\Bar{v}$ and $\Bar{m}$ is exactly three, as referred to in main text. For the \textbf{swap} part, one uses a $\braidtwo(\Bar{w})$ gate, which swaps positions of zero and one modes in $\Bar{m}$.

For the B subcolumn, we apply \textbf{shrink-and-swap} until we have 1 at pivot and pivot+1 position, and note that this does not affect any of the previous columns. Note that if B has no zero entries, \textbf{shrink} will not work, and we need an extra bit initialized to zero to have an overlap of three. For the A subcolumn, we use the pivot mode as a ``control" mode in the \textbf{shrink} part, if control mode is 1, then shrink happens, else it does not, so it only affects the last column. We perform this \textbf{shrink-and-swap} on A until all the entries are zero. The product of these $\braid$ and $\braidtwo$ gates corresponds to $L_{2n\times 2n}$:
\begin{align}
    L_{2n\times2n}S'_{2n\times k+1} = S^{d}_{2n\times k+1}
\end{align}and this concludes the proof.

\section{\\Details of $\braidtwo$ and $\braid$ calculation }

The \textbf{$\braidtwo$} operator is defined to be:
\begin{align}
\text{$\braidtwo$}(i,j) &= \exp(-\frac{\pi}{4} c_i c_j)\\
          &= \frac{1}{\sqrt{2}}(1 -  c_ic_j)
\end{align}
In our shorthand notation, we write the above as:
\begin{align}
\text{$\braidtwo$}(\bar{c}) = \frac{1}{\sqrt{2}}(1 - [\bar{c}])   
\end{align}
The conjugate action of this operator on a Majorana string $\bar{m}$ is:
\begin{multline}
\braidtwo(\bar{c})(\bar{m})\braidtwo(\bar{c})^{\dagger} = \frac{1}{2}(1-[\bar{c}][\bar{m}]\\ + [\bar{m}][\bar{c}] - [\bar{c}][\bar{m}][\bar{c}])
\end{multline}
Since $[\bar{m}][\bar{c}]=(-1)^{|\bar{c}\Lambda_f \bar{m}|} [\bar{c}][\bar{m}]$\\
\begin{multline}
\braidtwo(\bar{c})(\bar{m})\braidtwo(\bar{c})^{\dagger} =\frac{1}{2}(1+(-1)^{|\Bar{c}\Lambda_f \Bar{m}|})[\bar{m}]+\\
\frac{1}{2}(-1+(-1)^{|\Bar{c}\Lambda_f\Bar{m}|})[\Bar{c}][\Bar{m}]
\end{multline}
In a binary matrix form, $\braidtwo(\bar{c})$ can be written as:
\begin{align}
\braidtwo(\Bar{c})_{\alpha,\beta}=
\begin{cases}
1 & \alpha = \beta ,\alpha,\beta \notin \Bar{c}\\
1 & \alpha \neq \beta,\alpha,\beta \in \Bar{c}\\
0 & otherwise
\end{cases}
\end{align}
as an example for the $\braidtwo(\Bar{c})$ where $\Bar{c}=c_1c_2$ is:
\begin{align}
\braidtwo(c_1c_2) =
\begin{bmatrix}
0 & 1 \\
1 & 0 \\
\end{bmatrix}
\end{align}
The phase update rule is:
\begin{align}
    r\to r + |\bar{v}^T \Lambda_f \bar{m}|(1+ 2\bar{v}^T \Lambda_f^L \bar{m})
\end{align}
where $\Lambda_f^L$ is the lower triangular part of $\Lambda_f$. 
Similarly for the \textbf{$\braid$} gate:
\begin{align}
\braid(\bar{c}) = \frac{1}{\sqrt{2}}(1 + i [\bar{c}])
\end{align}
\begin{multline}
\braid(\bar{c})(\bar{m})\braid(\bar{c})^{\dagger} =\frac{1}{2}(1+(-1)^{|\Bar{c}\Lambda_f \Bar{m}|})[\bar{m}]+\\
\frac{i}{2}(1-(-1)^{|\Bar{c}\Lambda_f\Bar{m}|})[\Bar{c}][\Bar{m}]
\end{multline}
In the binary matrix notation, $\braid(\Bar{v})$ can be written as:
\begin{align}
\braid{(\Bar{v})}_{\alpha,\beta} = 
\begin{cases}
1 & \alpha=\beta, \alpha,\beta \notin \Bar{v} \\
1 & \alpha,\beta \in \Bar{v}, \alpha \neq \beta \\
0 & otherwise
\end{cases}
\end{align}
as an example, $\braid(\bar{v})$, where $\bar{v} = c_1c_2c_3c_4$ is:
\begin{align}
    \braid({c_1c_2c_3c_4}) = 
     \begin{bmatrix}
0 & 1 & 1 & 1 \\
1 & 0 & 1 & 1 \\
1 & 1 & 0 & 1 \\
1 & 1 & 1 & 0 \\
\end{bmatrix}  
\end{align}
The general derivation for conjugating by $\exp(i \frac{\pi}{4} V)$ is:
\begin{align}
    e^{i\frac{\pi}{4}(i^{r'}[\bar{v}])}&i^r[\bar{m}] e^{-i\frac{\pi}{4}(i^{r'}[\bar{v}])}  \\
    &= \frac{1}{2}(I + i(i^{r'}[\bar{v}])) i^r[\bar{m}] (I - i(i^{r'}[\bar{v}]))\\
    &= \frac{1}{2}(i^r[\bar{m}] + i^r (i^{r'}[\bar{v}])[\bar{m}](i^{r'}[\bar{v}]))\\
    &~~~~+\frac{1}{2}(i(i^{r'}[\bar{v}])(i^r[\bar{m}]) - i(i^r[\bar{m}]) (i^{r'}[\bar{v}]))\\
    &= \frac{1}{2}(1+(-1)^{\bar{v}^T\Lambda_f\bar{m}})i^r[\bar{m}] \\&~~~~+ \frac{1}{2}(1-(-1)^{\bar{v}^T\Lambda_f\bar{m}})i^{r+r'+1}[\bar{v}][\bar{m}] \\
    &= \frac{1}{2}(1+(-1)^{\bar{v}^T\Lambda_f\bar{m}})i^r[\bar{m}] \\&~~~~+ \frac{1}{2}(1-(-1)^{\bar{v}^T\Lambda_f\bar{m}})i^{r+r'+1 + 2\bar{v}^T\Lambda_f^L\bar{m}}[\bar{v}+\bar{m}]
\end{align}

\section{Proof there is no ancilla free method to decode if $P_{tot}$ is inside the stabilizer}
\begin{proof}
Suppose $P_{tot}$ is inside $S^{e}$. One can multiply the columns of $S^{e}$ such that one of the columns corresponds to all ones. Then acting on this column with $\braid$ or $\braidtwo$ does not change this column since the overlap is always even, so the parity stays inside the stabilizer. Therefore, it is not possible to obtain the decoded stabilizer and thus get a decoding circuit.  
\end{proof}

\end{document}